\documentclass[conference]{IEEEtran}
\usepackage[utf8]{inputenc}
\usepackage[T1]{fontenc}

\usepackage{caption}
\usepackage{subcaption} 

\usepackage[
  style=ieee,
  backend=bibtex, 
  bibencoding=ascii,
  hyperref=false,
  backref=false,
  doi=false,
  isbn=false,
  maxnames=1, 
  minnames=1,
  maxbibnames=999
]{biblatex}
\usepackage{csquotes} 
\bibliography{bibliography} 

\usepackage{amsmath}
\usepackage{amssymb}
\usepackage[binary-units,per-mode=symbol]{siunitx}
\usepackage{mathtools}

\usepackage{graphicx}
\graphicspath{{./image/}}

\usepackage{tikz}
\usepackage[absolute,overlay]{textpos}

\usetikzlibrary{arrows,shadows,decorations.pathmorphing,backgrounds,fit,positioning,calc,shapes}

\usepackage{pgfplots}
\pgfplotsset{compat=newest}
\newlength\figureheight
\newlength\figurewidth

\usepackage{ctable}
\usepackage{makecell} 

\usepackage{adjustbox}

\newcolumntype{R}[2]{%
    >{\adjustbox{angle=#1,lap=\width-(#2)}\bgroup}%
    l%
    <{\egroup}%
}
\newcommand*\roth{\multicolumn{1}{R{45}{1em}}}

\usepackage{stmaryrd}

\usepackage{xcolor}
\usepackage{soul}
\usepackage{amsmath}
\usepackage[binary-units,per-mode=symbol]{siunitx}
\usepackage{relsize}

\usepackage[normalem]{ulem}

\definecolor{tuBlue}{rgb}{0.298039215686275,0.603921568627451,0.713725490196078}
\definecolor{White}{rgb}{1,1,1}

\newcommand*{\rely}[2]{\setlength{\fboxsep}{0pt}\colorbox{tuBlue!#2!White}{#1}}
\newcommand*{\unrely}[2]{\setlength{\fboxsep}{0pt}\colorbox{tuBlue!#2!White}{\sout{#1}}}
\begin{document}
%
\title{BANDANA -- Body Area Network Device-to-device Authentication using Natural gAit}

\author{\IEEEauthorblockN{Dominik Schürmann\IEEEauthorrefmark{1},
Arne Brüsch\IEEEauthorrefmark{1}\IEEEauthorrefmark{2},
Stephan Sigg\IEEEauthorrefmark{2} and
Lars Wolf\IEEEauthorrefmark{1}}
\IEEEauthorblockA{\IEEEauthorrefmark{1}Institute of Operating Systems and Computer Networks, TU Braunschweig\\
Email: \{schuermann, bruesch, wolf\}@ibr.cs.tu-bs.de}
\IEEEauthorblockA{\IEEEauthorrefmark{2}Ambient Intelligence, Comnet, Aalto University\\
Email: \{arne.bruesch, stephan.sigg\}@aalto.fi}}


%


\IEEEspecialpapernotice{(Extended Version of PerCom 2017 Submission)}

\maketitle

\begin{abstract}
Secure spontaneous authentication between devices worn at arbitrary location on the same body is a challenging, yet unsolved problem. We propose BANDANA, the first-ever implicit secure device-to-device authentication scheme for devices worn on the same body. Our approach leverages instantaneous variation in acceleration patterns from gait sequences to extract always-fresh secure secrets. It enables secure spontaneous pairing of devices worn on the same body or interacted with. The method is robust against noise in sensor readings and active attackers. We demonstrate the robustness of BANDANA on two gait datasets and discuss the discriminability of intra- and inter-body cases, robustness to statistical bias, as well as possible attack scenarios.
\end{abstract}


%
\IEEEpeerreviewmaketitle

\section{Introduction}
Scalable secure device pairing in the wake of the Internet of Things (IoT) is a pending problem that has not yet been solved satisfactorily.
Current pairing protocols include pin-based approaches (e.g. Bluetooth) or out-of-band communication~\cite{sethi_bootstrapping_2014}.
Example out-of-band channels are, for instance, a secret printed on or displayed by a device, Near Field Communication~\cite{suomalainen_NFC_2007} or audio\footnote{A popular commercially implemented example using audio to initiate device pairing is the chromecast protocol}\cite{schurmann2013secure}. 
Alternatively, the standard approach for IoT security is to have the device connect to a dedicated trusted server which then handles the pairing and necessary key-exchange of the devices in question~\cite{sicari2015security}. 

These approaches are sufficient for one-time manual pairing of a limited number of devices. 
However, the personal device-network in the IoT is expected to experience frequent fluctuation in device count and identity~\cite{chong2014survey}.
New devices are added in the context of use while others are discarded. 
Examples are pairings to a multitude of body-worn smart devices (watches, glasses, smartphone, bio-sensors), fitness trackers, changing smart textile, or to temporarily used external devices such as bicycles, cars, shopping carts, or equipment in a fitness center.
While seamless pairing without manual user interaction among such devices promises new, personalised services, the threat of privacy exposion to malicious adversaries needs to be controlled by novel secure pairing schemes that scale. 

\begin{figure}
\centering
\includegraphics[width=1\columnwidth]{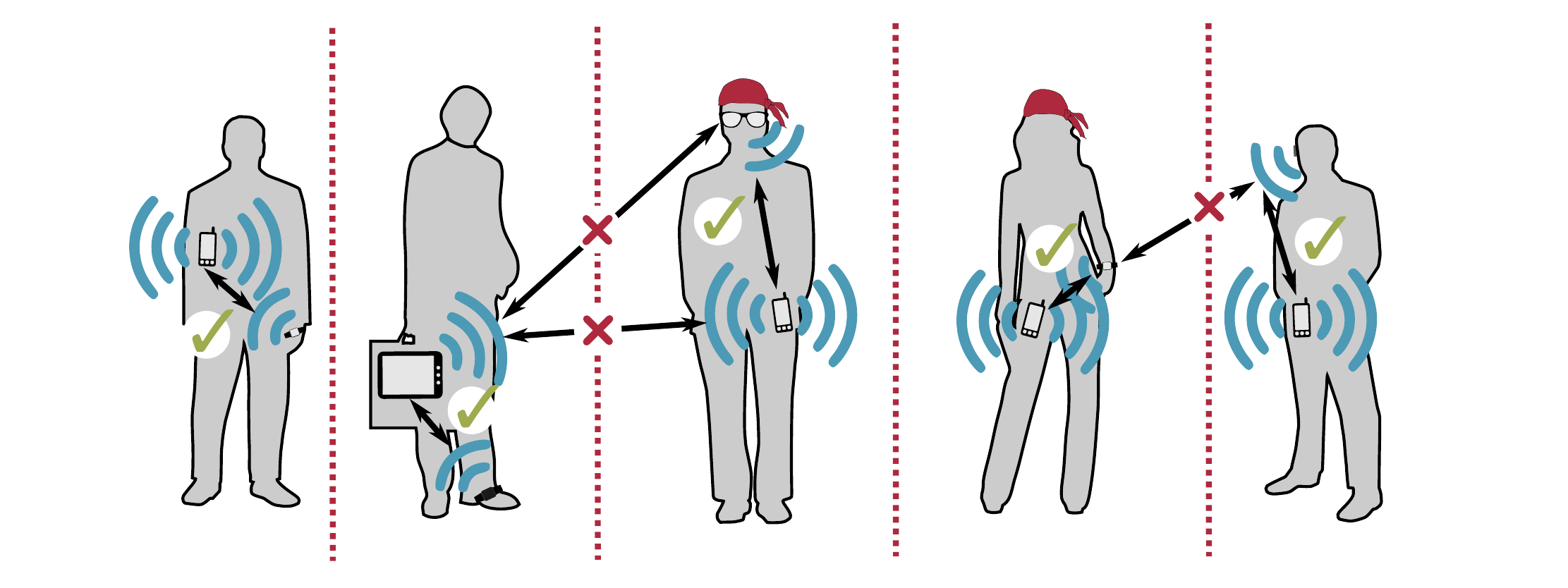}
\caption{BANDANA creates implicit security barriers towards devices in proximity, while establishing ad-hoc spontaneously secure connections between devices worn on the same body.}
\label{fig:scenario}
\end{figure}

As depicted in Figure~\ref{fig:scenario}, we envision spontaneous secure pairing which allows frequent re-pairing (restricted to the time-of-use), and ad-hoc implicit (no manual interaction required) secure authentication bound to an individual.
The desired solution shall not require a trusted third party.

This paper proposes a solution to this challenge by introducing a secure pairing scheme among on-body devices based on common movement patterns due to co-location on the same body. 
In particular, we exploit instantaneous variations in gait sequences for implicit generation of a shared secret among all devices on the same body. 
The contributions of our work are 
\begin{enumerate}
\item a secure ad-hoc pairing scheme for devices worn on the same body
\item the experimental verification of the protocol on a number of large-scale gait datasets
\item security analysis on the pairing approach covering Entropy, statistical bias, and attack scenarios
\end{enumerate}

The remainder of the paper is organized as follows. 
In section~\ref{sectionRelatedWork} we introduce previous studies on ad-hoc secure device pairing and how they relate to our work. 
We will then in section~\ref{sec:fundamentals} introduce basic methodology on gait cycle detection, datasets utilised in our evaluation, and data preparation, before introducing our secure ad-hoc pairing scheme, BANDANA in section~\ref{sec:pairing}.
In section~\ref{sec:attackmodel}, attack models on the protocol are discussed and in section~\ref{sec:evaluation}, the proposed ad-hoc pairing based on instantaneous gait-patterns is evaluated on a number of large-scale gait datasets and with respect to Entropy, statistical bias, as well as random and sophisticated attacks. 
Finally, section~\ref{sec:conclusion} concludes our discussion.


\section{Related Work}\label{sectionRelatedWork}

A popular sensor to detect co-presence is the accelerometer. 
For instance,~\cite{bichler2007key} present a process to generate shared keys based on shaking processes. 
They propose a threshold-based protocol conditioned on the magnitude of the co-aligned acceleration processes. 
A similar approach has been followed by Mayrhofer et al.~\cite{mayrhofer2007shake}, who demonstrated that an authentication is possible when devices are shaken simultaneously by a single person, while an authentication was unlikely for a third person trying to mimic the correct movement pattern remotely. 
Improvements to this protocol have been proposed, for instance, in~\cite{groza2012saphe}, where the success probability of devices in proximity is increased by considering acceleration patterns derived in the order of their magnitude. 
Also,~\cite{liu2014overlapped} solves the issue of different sample rates on paired devices and non-aligned starting points while~\cite{mayrhofer2015optimal} mitigate relative differences in rotation among devices.  
Based on this protocol,~\cite{findling2014shakeunlock,findling2016shakeunlock} presents an approach to unlock a mobile device with the help of a smartwatch when both are shaken simultaneously. 
Their approach, however, requires that 
acceleration sequences are exchanged and compared via an established secure channel. 
Another implementation was presented in~\cite{van2016accelerometer} to authenticate from the acceleration pattern of a vibrating device lying on a flat surface. 
Applications for such schemes have been proposed, for instance, in~\cite{gu2014toauth} where devices connected via NFC 
are authenticated via acceleration traces of vibrating smart phones that touch each other.
Also,~\cite{mehrnezhad2014tap} 
propose to prevent man-in-the-middle relay attacks on NFC payment 
by requiring matching acceleration sequences generated during double-tapping one device on the other~\cite{mehrnezhad2015tap}

For authentication based on arbitrary co-aligned sensor data, the candidate key protocol is proposed in~\cite{mayrhofer2007candidate}. 
It interactively exchanges hashes from feature sequences as short secrets and concatenates the key from the secrets with matching hashes. 

Examples for possible sensor modalities that can be used for unattended co-presence-based device pairing apart of acceleration are magnetometer~\cite{jin2016magpairing}, RF-signals~\cite{varshavsky2007amigo,knox2015wireless} luminosity~\cite{miettinen2014context} or audio~\cite{schurmann2013secure}.
These, however, have in common that the pairing is not constrained to devices on the same body but, more generally, to devices in proximity.

We are, in contrast, focusing on the pairing of devices on the same body or interacted with by the same person. 
Consider, for instance,~\cite{shu2014dynamic}, where key-card authentication is extended with additional acceleration patterns.
Another example is presented in~\cite{mare2014zebra} to authenticate a user wearing an acceleration-capturing bracelet during interaction with a keyboard device. 
The device is locked when keyboard-interaction and acceleration patterns mismatch. 

Very related to our study is the work of Cornelius et al.~\cite{cornelius2011recognizing} to identify devices co-located on the same body via correlated acceleration readings.
An important problem to consider when comparing acceleration sequences among devices on the body is that orientation and placement significantly impact the recorded acceleration, gyro and magnetometer sequences~\cite{kunze2011compensating}. 
Solutions to receive placement independent features are, (A) to calculate the norm or magnitude $m_i = \sqrt{x^2_i + y^2_i+ z^2_i}$ (thereby discarding information on acceleration along individual axes~\cite{muaaz2014orientation}), (B) to first detect the location on the body and then to try to deal with changes that occur due to placement~\cite{kunze2011compensating}, or (C) to tackle disorientation and misplacement errors by calculating the rotation matrix from magnetometer readings~\cite{hoang2013lightweight}. 
Even though after (A), the resulting signal still differed greatly due to inherently differing movement of underlying body parts (e.g. arm vs. head vs. legs)~\cite{heinz2003experimental}, Cornelius et al.~\cite{cornelius2011recognizing} succeeded to show good correlation among all body locations from mean, standard deviation, variance, mean absolute deviation and interquartile range as well as signal's energy.
This result is a strong indication that secure keys conditioned on co-location on the same body exist.
However, as correlation can be alternating positively and negatively, it remains unsolved how this can be exploited for the generation of keys, when the sequences shall not be disclosed to an adversary listening to any communication between nodes.

A weakness of using arbitrary acceleration sequences for spontaneous pairing of on-body devices is that for a significant number of daily activities, upper body and lower body movements are only weakly or not correlated.
An activity that can be well recognized over the whole body though is walking or gait~\cite{muaaz2013analysis}.
For instance, identical step patterns from acceleration sequences have been utilised for co-location detection~\cite{srivastava2015step}.
The authors in~\cite{hoang2015gaitauthentication,hoang2013leightweight} employ gait cycles to authenticate a user on his smart-phone by
matching the current walking pattern against a previously saved walking template exploiting a fuzzy commitment scheme~\cite{juels1999fuzzy}.
Another gait-based authentication system is proposed in~\cite{casale2012personalization} where a two-stage classifier first distinguishes walking from other activity to then exploit individual gait patterns. 
A concern for any authentication scheme based on implicit features is that the generated key sequences employ a high randomness and are uniformly distributed over the key space so that an adversary can not easily guess or re-produce the key. 
To this end, \cite{hoang2015gait} recently presented an approach to generate a key fingerprint from the difference of a mean world gait (spanning the complete population) to the mean gait of an individual. 
By computing the mean gait over the whole population, the authors assured that the resulting sequence is well balanced and uniformly distributed.

\begin{figure*}
  \centering
  \begin{subfigure}[t]{\columnwidth}
    \centering%
    \scriptsize
    \setlength\figureheight{3.5cm}
    \setlength\figurewidth{\columnwidth}
    \input{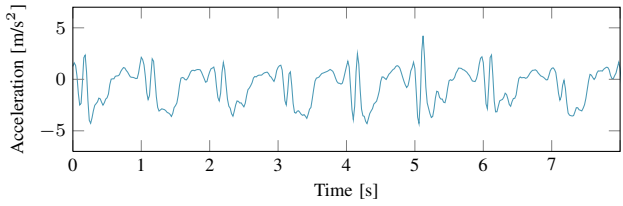}
    \caption{Unmodified accelerometer reading (z-axis) at \SI{50}{\hertz}.}
    \label{fig:algo1}
  \end{subfigure}%
  \quad
  \begin{subfigure}[t]{\columnwidth}
    \centering%
    \scriptsize
    \setlength\figureheight{3.5cm}
    \setlength\figurewidth{\columnwidth}
    \input{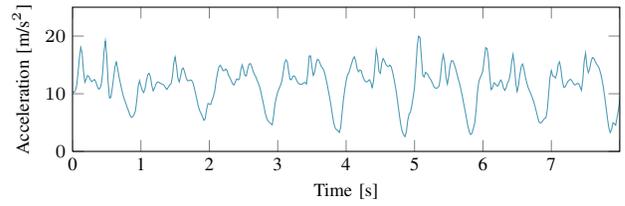}
    \caption{Application of Madgwick's algorithm. It should be noted that the gravity $g = {\sim\SI{9.81}{\meter\per\second\squared}}$ can now be recognized, indicating a correct orientation relative to the ground.}
    \label{fig:algo2}
  \end{subfigure}
  \par\smallskip
  \begin{subfigure}[t]{\columnwidth}
    \centering%
    \scriptsize
    \setlength\figureheight{3.5cm}
    \setlength\figurewidth{\columnwidth}
    \input{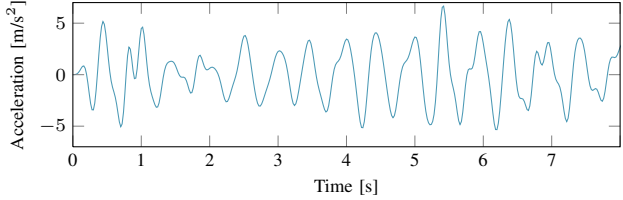}
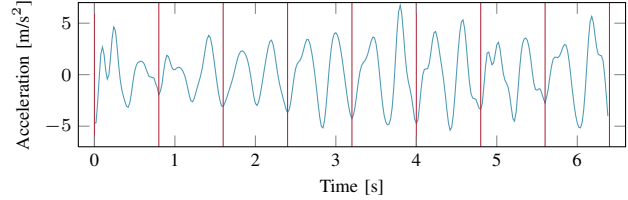
    \caption{Application of Type-II Chebyshev bandpass filter.}
    \label{fig:algo3}
  \end{subfigure}%
  \quad
  \begin{subfigure}[t]{\columnwidth}
    \centering%
    \scriptsize
    \setlength\figureheight{3.5cm}
    \setlength\figurewidth{\columnwidth}
\begin{tikzpicture}

\definecolor{color1}{rgb}{0.682352941176471,0.16078431372549,0.243137254901961}
\definecolor{color0}{rgb}{0.298039215686275,0.603921568627451,0.713725490196078}

\begin{axis}[
xlabel={Time [\si{\second}]},
ylabel={Acceleration [\si{\meter\per\second\squared}]},
xmin=-10, xmax=330,
ymin=-7, ymax=7,
axis on top,
width=\figurewidth,
height=\figureheight,
xtick={0,50,100,150,200,250,300},
xticklabels={0,1,2,3,4,5,6}
]
\addplot [color0]
table {%
0 -4.6980726423317
1 -4.68003891689398
2 -2.65548187222276
3 -0.206551697685933
4 2.01012119029231
5 2.6811922776366
6 1.94689857044404
7 0.441661544856334
8 -0.430318198877477
9 -0.0108917301469119
10 1.66738620453246
11 3.54787526380019
12 4.62580204985543
13 4.28927257110999
14 2.86099853024148
15 1.03757040804071
16 -0.464225052597854
17 -1.45725582759578
18 -2.09618678862292
19 -2.63790415174675
20 -3.07182949273978
21 -3.16838689542268
22 -2.71904741641491
23 -1.76156733058889
24 -0.630671861144616
25 0.331718436371324
26 0.90223396397134
27 1.17495685418263
28 1.25894859857944
29 1.31494600164781
30 1.23748322758979
31 1.03215061843569
32 0.596057742937712
33 0.179589079944651
34 -0.176579567806352
35 -0.222744039106403
36 -0.276111309761687
37 -0.313506873235936
38 -0.835624196274726
39 -1.238207278837
40 -1.93291669344737
41 -1.60679497027555
42 -0.896528519003779
43 0.317119348253377
44 1.33728532443971
45 1.87644945558942
46 1.74785180935138
47 1.26985411872888
48 0.749075662583812
49 0.49315473354671
50 0.486368760855025
51 0.626166601539385
52 0.716557307373888
53 0.703385119383851
54 0.564762444385943
55 0.33159871775205
56 -0.0453455575762578
57 -0.602230830172446
58 -1.32449045559027
59 -2.04916487842278
60 -2.54414071100051
61 -2.63058632133397
62 -2.30918440499713
63 -1.75657887552082
64 -1.17288332291108
65 -0.650448192348829
66 -0.0922812359536903
67 0.634687454155782
68 1.59570961722951
69 2.62333132428029
70 3.46088103570617
71 3.81010527873074
72 3.59337377713267
73 2.84136810070853
74 1.78943204191636
75 0.569002879962608
76 -0.60411829559656
77 -1.70704001038426
78 -2.49788616712357
79 -3.03255398831615
80 -3.0865081338553
81 -2.76341373289277
82 -2.33939154280161
83 -1.82707871746601
84 -1.35765935764053
85 -0.864672957757291
86 -0.326236108917036
87 0.312486239511246
88 0.981661467243758
89 1.59863886544939
90 2.04327511035716
91 2.27838439237496
92 2.30472941450629
93 2.1775797230579
94 1.91215665964506
95 1.48812986231475
96 0.840440855755663
97 -0.0486490436287301
98 -1.09795431528308
99 -2.0978727331947
100 -2.7906582514862
101 -2.9864014248584
102 -2.65674691569406
103 -1.93411745828772
104 -1.00963782684895
105 -0.0354386231907797
106 0.931481723420305
107 1.85776889831741
108 2.6774457263937
109 3.22878362377756
110 3.34914277227621
111 2.93799710851313
112 2.08662791608822
113 1.00038612223265
114 -0.0392794186881201
115 -0.896809812490203
116 -1.54791717453776
117 -2.15138012286057
118 -2.74365278307164
119 -3.33719698801363
120 -3.67357901357179
121 -3.39984052517896
122 -2.64700363165688
123 -1.46456456646449
124 -0.293974913434129
125 0.591924176903629
126 1.05121457919423
127 1.26030896184357
128 1.45852752375201
129 1.86010950576581
130 2.44692646877511
131 3.052447011855
132 3.42099139572947
133 3.39698600360707
134 2.94729333045531
135 2.16506883549788
136 1.1473976099022
137 -0.0458066888070457
138 -1.3972407072828
139 -2.83083707106419
140 -4.14586357460166
141 -5.02501869303303
142 -5.16113104207452
143 -4.42082338839517
144 -2.92871492708896
145 -1.03488683987651
146 0.848221288325368
147 2.38521209497614
148 3.42723322278122
149 3.9637260545441
150 4.07815711493996
151 3.82848918486472
152 3.27395061260015
153 2.4393176170426
154 1.41220071655187
155 0.254733161340708
156 -0.923569617691054
157 -2.0915443205245
158 -3.10749345073063
159 -3.89259526826915
160 -4.23383977945706
161 -3.86774349873668
162 -3.0610891711488
163 -1.85623939042893
164 -0.663373467728991
165 0.30385007575325
166 0.878067885767558
167 1.17168008212586
168 1.34874334532938
169 1.63387552926654
170 2.0981001322954
171 2.71108587265436
172 3.29255658877317
173 3.64696623896627
174 3.57486809191766
175 2.9734166716822
176 1.83517913181193
177 0.305412693564316
178 -1.35955194541109
179 -2.85155468171354
180 -3.94842093825738
181 -4.59238395893319
182 -4.85804631629372
183 -4.80480119729871
184 -4.32609823554696
185 -3.1837675115823
186 -1.20209310053056
187 1.44090872373315
188 4.15917908839074
189 6.12526492254577
190 6.72504461927256
191 5.84642582222016
192 3.99936305428326
193 1.94823865973725
194 0.355872845441895
195 -0.649118052141914
196 -1.32621194582554
197 -2.15531863845017
198 -3.22652446295955
199 -4.34376747538206
200 -4.71248372044447
201 -4.42038272383614
202 -2.98316172299064
203 -1.34697061510296
204 -0.0211005943158503
205 0.516154236512395
206 0.56070418156201
207 0.57605506575144
208 1.09202101524712
209 2.10904015453156
210 3.29821695479655
211 4.09094262417698
212 4.18492320587115
213 3.58363014055951
214 2.61301751306898
215 1.55957127212499
216 0.53604091530564
217 -0.588081079506327
218 -1.93480278508983
219 -3.43296365860732
220 -4.73288558906349
221 -5.38706677346759
222 -5.09938544026811
223 -3.87005549524492
224 -1.97487123349507
225 0.218199207126128
226 2.33396285773386
227 4.07542047169043
228 5.13735486993488
229 5.31442830618894
230 4.50163585208617
231 2.91720816706236
232 0.968323218622438
233 -0.743084215467115
234 -1.8629020392302
235 -2.2647499939987
236 -2.32833417589403
237 -2.38773737389164
238 -2.82105599572393
239 -3.26434602250383
240 -3.35688615395139
241 -2.78551684854952
242 -1.05689180100839
243 0.821521495944254
244 2.09275813402989
245 2.19252824807381
246 1.55653851957528
247 0.970543358903598
248 1.14608856317308
249 1.99919854610567
250 2.91488105630528
251 3.14520095726588
252 2.48502271958503
253 1.2751935228715
254 0.151722135035757
255 -0.5202274308449
256 -0.824571209816572
257 -1.20420284619596
258 -2.01565926433131
259 -3.20919707387028
260 -4.26239959566047
261 -4.51499062810071
262 -3.62569379164644
263 -1.82016041136206
264 0.249271774068161
265 1.95349582844837
266 2.99717989312175
267 3.47468610697751
268 3.55786399647785
269 3.32450825869214
270 2.67382660484434
271 1.61363544037154
272 0.314745880562552
273 -0.798332428902583
274 -1.4361175983309
275 -1.50709954482466
276 -1.38091994710039
277 -1.39298572691287
278 -1.83866155676496
279 -2.33128849486092
280 -2.74862126547462
281 -2.14094186440473
282 -1.35307323049795
283 -0.156042551273531
284 0.817074543180313
285 1.49671796105491
286 1.69702735272701
287 1.69822305238761
288 1.65344166829538
289 1.82777871609558
290 2.18240404072791
291 2.64191186265429
292 2.92934568587326
293 2.86346802717586
294 2.28082369585229
295 1.23062357098066
296 -0.14972894699416
297 -1.60023305841413
298 -2.9126053100017
299 -3.94739984251152
300 -4.68139330997758
301 -5.10998344903533
302 -5.15656635199158
303 -4.65137173374542
304 -3.39597939678669
305 -1.38556878389767
306 1.11230308992869
307 3.4976821705991
308 5.15635708303289
309 5.64476619407523
310 5.05459982958227
311 3.82140867383892
312 2.66585866344369
313 1.98791623938215
314 1.88675464807021
315 1.83462704952674
316 1.36142142844104
317 -0.0280056363073855
318 -1.94882974528206
319 -4.02020232973521
};
\path [draw=color1] (axis cs:0,-7)
--(axis cs:0,7);

\path [draw=color1] (axis cs:40,-7)
--(axis cs:40,7);

\path [draw=color1] (axis cs:80,-7)
--(axis cs:80,7);

\path [draw=color1] (axis cs:120,-7)
--(axis cs:120,7);

\path [draw=color1] (axis cs:160,-7)
--(axis cs:160,7);

\path [draw=color1] (axis cs:200,-7)
--(axis cs:200,7);

\path [draw=color1] (axis cs:240,-7)
--(axis cs:240,7);

\path [draw=color1] (axis cs:280,-7)
--(axis cs:280,7);

\path [draw=color1] (axis cs:320,-7)
--(axis cs:320,7);

\end{axis}

\end{tikzpicture}
    \caption{Resampling to $\rho=40$ and Gait Cycle Detection with $q=8$ gait cycles.}
    \label{fig:algo4}
  \end{subfigure}
  \caption{Example of our process for sensor data pre-processing and gait cycle detection. Here, the z-axis of an accelerometer is depicted, which is attached to the forearm of one subject.}
  \label{fig:algo}
\end{figure*}

These studies on gait-based authentication (1) do not address the impact of different on-body locations and sensor orientation and (2) intend to use gait as a unique biometric feature that does not change for an individual over time. 
In contrast, in our case, we intend to generate an always-fresh authentication key based on instantaneous acceleration sequences for arbitrary location on the human body. 
Muaaz et al.~\cite{muaaz2015cross} confirmed the significant challenge of (1) but demonstrated gait-based authentication covering closely related locations on the human body (from one to the other side of the hip) is possible but suffers from high error rate. 
 
For the verification and security analysis of gait-based authentication schemes, it is crucial to test the approach on a large number of participants. 
To this end, \cite{gafurov2007spoof} present a gait dataset for 100 participants wearing accelerometers at the hip. 
Another dataset with 744 subjects has been presented in~\cite{ngo2014largest}.
Participants traverse a parcours featuring distinct walking conditions such as straight, coarse ground, upwards, downwards and steps.
Also, Sztyler et al. created a real-world dataset used for Position Aware Activity Recognition~\cite{sztyler2016onbody} in which 15 subjects, equipped with sensors on 7 distinct locations on the body performed different actions for a time period of approximately 10--12 minutes each. 

Another, conceptional challenge with all context-based authentication approaches is that due to sensing inaccuracies, different hardware and noise the sensed signals are likely not identical but only similar. 
Fuzzy cryptography presents a methodology to obtain identical keys 
from similar patterns~\cite{juels1999fuzzy}. 
In particular, by mapping the patterns into the codespace of an error correcting code, mismatches can be mitigated without disclosing the pattern over a potentially insecure channel. 

\section{Fundamentals}\label{sec:fundamentals}

\subsection{Gait Cycle Detection}
\label{sec:gaitcycle}

In this section, our gait cycle detection algorithm is presented which builds on ideas by \citeauthor{hoang2015gaitauthentication}\cite{hoang2015gaitauthentication,hoang2013leightweight}.
In addition, we also utilize gyroscope readings to normalize the sensor's orientation and keep only the z-Axis that points in the opposite direction of gravity.

\begin{figure}
\centering
\includegraphics[width=1\columnwidth]{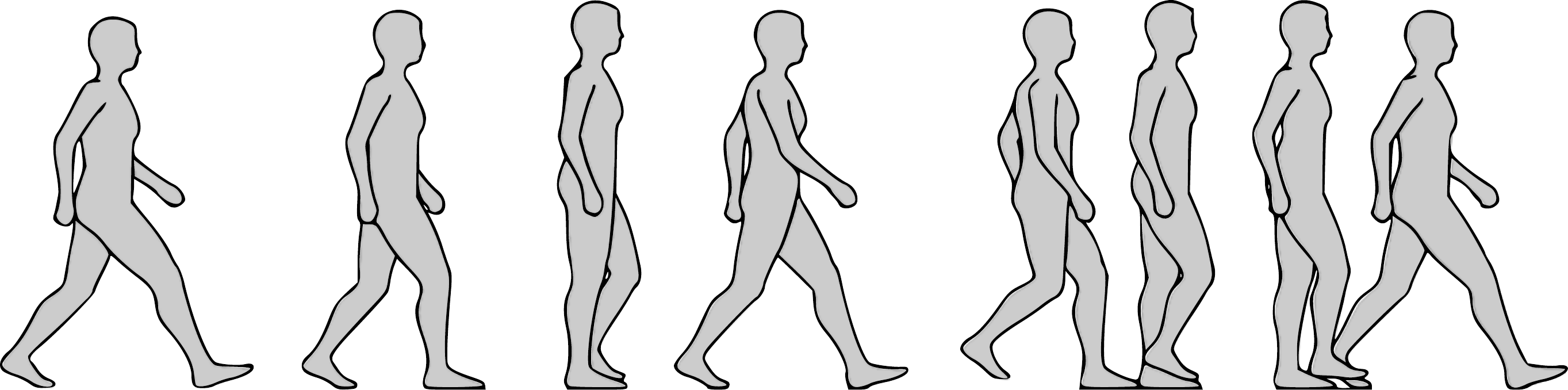}
\caption{A full gait cycle starting with the initial contact of the right foot. Characteristic intermediate positions are depicted separately. (modified from \cite{simoneau2002kinesiology})}
\label{fig:gait}
\end{figure}
As depicted in Figure~\ref{fig:gait}, a gait cycle is defined as the \enquote{time interval between two successive steps
}
, e.g., if a cycle starts with the initial contact of the right foot, it ends with the same position of the right foot~\cite{whittle2007normalgait}.
Thus, the goal of the algorithm is to find repetitive cycles in the raw acceleration signal.
Its input is a vector of amplitude values
\begin{equation*}
\boldsymbol{z} = \boldsymbol{(}z_1,\dots, z_n\boldsymbol{)}
\end{equation*}
of the accelerometer z-axis (cf. Figure~\ref{fig:algo1}).
Its output is a gait sequence of consecutive gait cycles with normalized length.




To find repetitive parts in the signal, a naive approach would search for local minima and then use these to split the raw signal into cycles.
However, the list of all local minima (comparing only direct neighbors) contains too many minima within single gait cycles.
Thus, this list must be filtered for minima having the same distance to each other to define clearly separated cycles.

Our filtering method is based on autocorrelation and distance calculation.
The discrete autocorrelation 
at time lag~$k$ and with variance $\sigma^2$ is estimated as 
\begin{equation*} 
\mathit{Acorr}(k) = \frac{1}{(n-k) \sigma^2} \sum_{t \in \mathbb{Z}} z_{t+k} \cdot \overline{z}_t
\end{equation*}
where $\overline{z}_t$ represents the conjugate of $z_t$.
The resulting autocorrelation
\begin{equation*}
\boldsymbol{a} = \boldsymbol{(}a_1,\dots, a_n\boldsymbol{)}
\end{equation*}
leads to $m$ non-ambiguous local maxima in $\boldsymbol{a}$, 
stored as
\begin{equation*}
\boldsymbol{\zeta} = \boldsymbol{\{}\zeta_1, \dots, \zeta_i, \dots \zeta_{m}\boldsymbol{\}} \text{.}
\end{equation*}
The distances between these indices and a mean distance
\begin{equation*}
\mathit{\delta_{mean}} = \left\lceil \frac{\sum_{i=1}^{m-1} \zeta_{i+1} - \zeta_i}{m-1} \right\rceil
\end{equation*} are calculated. 
$\mathit{\delta_{mean}}$ defines the length of \emph{half} a cycle, i.e., the time between the initial contact of the starting foot followed by the initial contact of the subsequent foot.
Thus, for $q$ describing the number of gait cycles, $m = q \cdot 2$.
For the gait-cycle extraction, we assume healthy subjects, where the movement of the right foot is sufficiently similar to the left foot and thus have nearly the same distance.
$\mathit{\delta_{mean}}$ can now be used to select indices of minima from $\boldsymbol{z}$ that represent clear cycles with the same length:
\begin{eqnarray}
\boldsymbol{\mu} &=& \boldsymbol{\{}\mu_1, \dots, \mu_{i}, \dots, \mu_{m-1}\boldsymbol{\}};\nonumber\\
\mu_i &=& \mbox{arg}\min (z_{\zeta_i-\tau}, z_{\zeta_i-\tau+1}, \dots, z_{\zeta_i+\mathit{\delta_{mean}}+\tau})\text{.}\nonumber
\end{eqnarray}
Every $\mu_j$ represents the index of a minimum in $\boldsymbol{z}$ limited to the range of $\mathit{\delta_{mean}}$ 
where 
$\tau$ defines an additional user defined factor to account for small deviations in the gait duration.
The indices in $\boldsymbol{\mu}$ can now be used to split the raw data $\boldsymbol{z}$ into full gait cycles
\begin{eqnarray}
\boldsymbol{Z} &=& \boldsymbol{\{}Z_1, \dots, Z_i, \dots, Z_q\boldsymbol{\}}; \nonumber\\
Z_i &=& \boldsymbol{(}z_{\mu_{\frac{i}{2}}}, \dots, z_{\mu_{i}}, \dots, z_{\mu_{\frac{i+1}{2}}-1}\boldsymbol{)};\label{equatonZi}\\
i&=&\{2,4,...,q\}\text{.}\nonumber
\end{eqnarray}
Finally, the length of gait cycles are normalized by resampling every $Z_i$ using a Fourier method to a fixed number of samples $\rho$ per gait cycle so that $|Z_i|=\rho$ (cf. Figure~\ref{fig:algo4}).
For ease of presentation, we will, in the following, describe such normalized gait cycle with $Z_i=\{Z_{i1},\dots,Z_{i\rho}\}$.
The choice of $\rho$ depends on factors such as sample rate and requirements of the quantization algorithm discussed in Section~\ref{sec:pairing}.

\subsection{Datasets}
\label{sec:datasets}
In our work, we used two different datasets gathered for evaluation of human movement (cf. Figure~\ref{figureDatasets}).
\begin{figure}
\fbox{\begin{minipage}{.98\columnwidth}
\begin{small}
\noindent\fcolorbox{gray!20}{gray!20}{\begin{minipage}{.96\columnwidth}\underline{\textit{Mannheim:}} The Real-World Dataset by Sztyler et al. was used for position aware activity recognition~\cite{sztyler2016onbody}. 
15 Subjects performed different actions for a time period of approximately 10 - 12 minutes each. 
They were equipped with 7 sensors on different parts of the body. 
These parts were chosen in order to gather data from every part of the body that acts different during human action.
\end{minipage}
}
\vspace{.3cm}

\noindent\fcolorbox{gray!20}{gray!20}{\begin{minipage}{.97\columnwidth}\underline{\textit{Osaka:}} The OU-ISIR Gait Database \cite{ngo2014largest} consists of samples taken from three triaxial accelerometers and gyroscopes worn on different parts of the thigh (left, right, center).
With this setup, subjects were asked to walk down a path, upstairs and down a slope. A total of 460 records exist within this dataset.
\end{minipage}}
\end{small}
\end{minipage}}
\caption{Datasets utilised in our study}
\label{figureDatasets}
\end{figure}
We refer to the datasets as the \textit{osaka} respectively the \textit{mannheim} dataset in accordance with the hometown of their collectors.
Unfortunately, the osaka dataset has certain faults.
A conceptual issue lies in the fact that all sensor units were located on pretty close positions on the body. As well, they were mounted to the same harness, introducing a possible error. 
Finally, 
the osaka dataset yields only 6-8 Gait Cycles of stationary walk per subject.


\subsection{Data Pre-Processing}
\label{sec:data_prep}
\begin{figure}
  \centering
  \begin{subfigure}[t]{0.30\columnwidth}
    \centering%
        \includegraphics[width=\columnwidth]{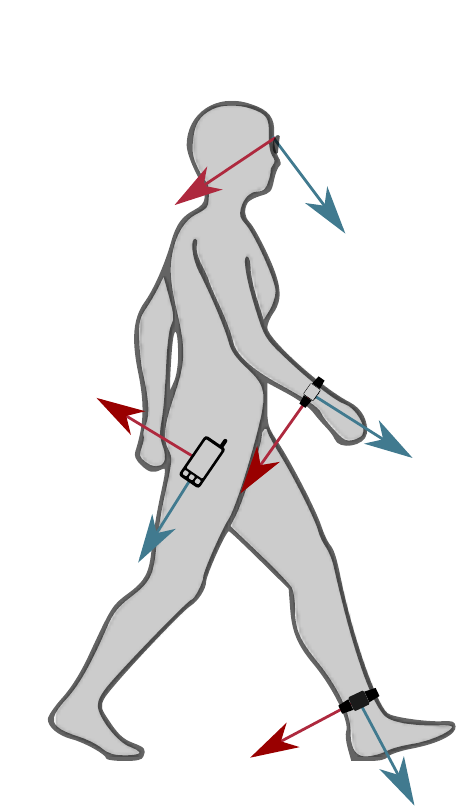}
        \caption{Before}
        \label{fig:orientation1}
  \end{subfigure}%
  \begin{subfigure}[t]{0.70\columnwidth}
    \centering%
    \def\svgwidth{0.8\columnwidth}
    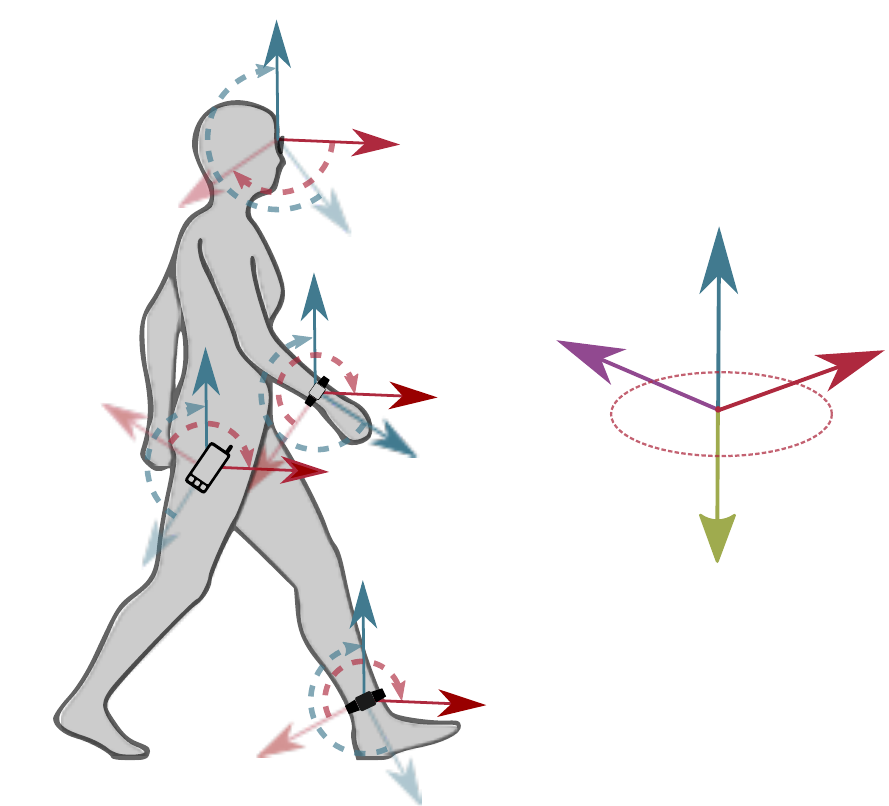
    \caption{After Madgwicks algorithm}
    \label{fig:orientation2}
  \end{subfigure}
  \caption{Different Sensors worn on a human body. Depicted are the devices' coordinate systems before and after application of Madgwicks algorithms. Note the remaining degree of freedom along the xy-plane.}
  \label{fig:orientation}
\end{figure}

In a real-world setting, it is very likely that different sensor platforms are worn at different positions, which introduce dynamically changing orientations due to body part movements (cf. Figure~\ref{fig:orientation1}).  
For our gait cycle detection to work well, it is crucial to align these every data point of these different orientations such that one of the axes is facing in the opposite direction of gravity as depicted in Figure~\ref{fig:orientation2}.

Nowadays, most mobile devices contain gyroscopes in addition to accelerometers~\cite{gsmarena2016gyros}.
We therefore posses information about the initial device orientation (since the force of gravity is included in every measurement recorded by the accelerometer) as well as the angular velocity of the sensor platform itself.
Thus, it is possible to correct the ongoing orientation error. We employ the algorithm proposed by Madgwick et al. \cite{madgwick2011estimation} to rotate all measurements $z_i$ accordingly, resulting in a signal as shown in Figure~\ref{fig:algo2}.
Note that the output is only guaranteed to be aligned along the z-axis which is in parallel with the gravity axis.
When comparing two readings, both other axes may point in different directions because no second fixed direction as, for example, the direction of North is obtainable.

Madgwick's algorithm only changes the sensor orientation (see Figure~\ref{fig:orientation}), it does not remove any noise.
Therefore, we apply a Type II
Chebyshev bandpass filter with passband chosen between \SI{0.5}{\hertz} and \SI{12}{\hertz}. Cutoff frequencies are further discussed in section~\ref{sec:bandpass}.
The resulting signal is shown in Figure~\ref{fig:algo3}.

\begin{figure*}
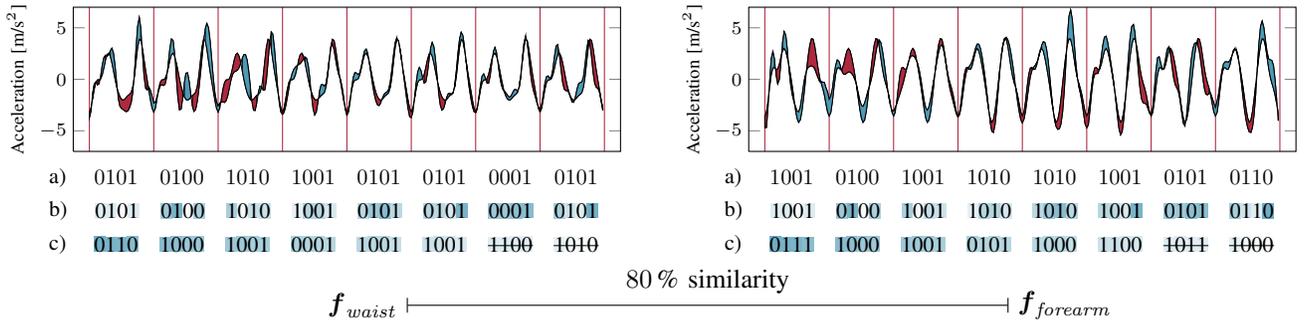

    \centering
    \begin{subfigure}[t]{\columnwidth}
    \begin{subfigure}[t]{\columnwidth}
        \centering%
        \scriptsize
        \setlength\figureheight{3.5cm}
        \setlength\figurewidth{\columnwidth}
        \input{image/algo5_waist_custom.tex}
    \end{subfigure}%
    \quad
    \begin{subfigure}[t]{\columnwidth}
        \resizebox {\columnwidth} {!} {
        \begin{tikzpicture}
          \node (0) at (0,0) {};
          \node[right=.6cm of 0] (1) {a)};
          \node[right=.2cm of 1] (2) {$0101$};
          \node[right=.1cm of 2] (3) {$0100$};
          \node[right=.1cm of 3] (4) {$1010$};
          \node[right=.1cm of 4] (5) {$1001$};
          \node[right=.1cm of 5] (6) {$0101$};
          \node[right=.1cm of 6] (7) {$0101$};
          \node[right=.1cm of 7] (8) {$0001$};
          \node[right=.1cm of 8] (9) {$0101$};
          \node[right=.3cm of 9] (10) {};
        \end{tikzpicture} 
        }
    \end{subfigure}
    \begin{subfigure}[t]{\columnwidth}
        \resizebox {\columnwidth} {!} {
        \begin{tikzpicture}
            \node (0) at (0,0) {};
            \node[right=.6cm of 0] (1) {b)};
            \node[right=.2cm of 1] (2) {\rely{0}{17}\rely{1}{18}\rely{0}{3}\rely{1}{18} };
            \node[right=.1cm of 2] (3) {\rely{0}{53}\rely{1}{75}\rely{0}{4}\rely{0}{46} };
            \node[right=.1cm of 3] (4) {\rely{1}{48}\rely{0}{6}\rely{1}{32}\rely{0}{29} };
            \node[right=.1cm of 4] (5) {\rely{1}{14}\rely{0}{41}\rely{0}{42}\rely{1}{23}};
            \node[right=.1cm of 5] (6) {\rely{0}{38}\rely{1}{55}\rely{0}{66}\rely{1}{38}};
            \node[right=.1cm of 6] (7) {\rely{0}{40}\rely{1}{38}\rely{0}{22}\rely{1}{74}};
            \node[right=.1cm of 7] (8) {\rely{0}{54}\rely{0}{54}\rely{0}{63}\rely{1}{66}};
            \node[right=.1cm of 8] (9) {\rely{0}{49}\rely{1}{28}\rely{0}{17}\rely{1}{80}};
            \node[right=.3cm of 9] (10) {};
        \end{tikzpicture}
        }

    \end{subfigure}
    \begin{subfigure}[t]{\columnwidth}
    \resizebox {\columnwidth} {!} {
    \begin{tikzpicture}
      \node (0) at (0,0) {};
      \node[right=.6cm of 0] (1) {c)};
      \node[right=.2cm of 1] (2) {\rely{0}{80}\rely{1}{75}\rely{1}{74}\rely{0}{66}};
      \node[right=.1cm of 2] (3) {\rely{1}{66}\rely{0}{63}\rely{0}{55}\rely{0}{54}};
      \node[right=.1cm of 3] (4) {\rely{1}{54}\rely{0}{53}\rely{0}{49}\rely{1}{48}};
      \node[right=.1cm of 4] (5) {\rely{0}{46}\rely{0}{42}\rely{0}{41}\rely{1}{40}};
      \node[right=.1cm of 5] (6) {\rely{1}{38}\rely{0}{38}\rely{0}{38}\rely{1}{32}};
      \node[right=.1cm of 6] (7) {\rely{1}{29}\rely{0}{28}\rely{0}{23}\rely{1}{22}};
      \node[right=.1cm of 7] (8) {\unrely{1}{18}\unrely{1}{18}\unrely{0}{17}\unrely{0}{17}};
      \node[right=.1cm of 8] (9) {\unrely{1}{14}\unrely{0}{6}\unrely{1}{4}\unrely{0}{3}};
      \node[right=.3cm of 9] (10) {};
    \end{tikzpicture}
    }
    \end{subfigure}
    \end{subfigure}
    \begin{subfigure}[t]{\columnwidth}
    \begin{subfigure}[t]{\columnwidth}
        \centering%
        \scriptsize
        \setlength\figureheight{3.5cm}
        \setlength\figurewidth{\columnwidth}
        \input{image/algo5_custom.tex}
    \end{subfigure}%
    \quad
    \begin{subfigure}[t]{\columnwidth}
        \resizebox {\columnwidth} {!} {
        \begin{tikzpicture}
          \node (0) at (0,0) {};
          \node[right=.6cm of 0] (1) {a)};
          \node[right=.2cm of 1] (2) {$1001$};
          \node[right=.1cm of 2] (3) {$0100$};
          \node[right=.1cm of 3] (4) {$1001$};
          \node[right=.1cm of 4] (5) {$1010$};
          \node[right=.1cm of 5] (6) {$1010$};
          \node[right=.1cm of 6] (7) {$1001$};
          \node[right=.1cm of 7] (8) {$0101$};
          \node[right=.1cm of 8] (9) {$0110$};
          \node[right=.3cm of 9] (10) {};
        \end{tikzpicture} 
        }
    \end{subfigure}
    \begin{subfigure}[t]{\columnwidth}
        \resizebox {\columnwidth} {!} {
        \begin{tikzpicture}
            \node (0) at (0,0) {};
            \node[right=.6cm of 0] (1) {b)};
            \node[right=.2cm of 1] (2) {\rely{1}{17}\rely{0}{18}\rely{0}{3}\rely{1}{18} };
            \node[right=.1cm of 2] (3) {\rely{0}{53}\rely{1}{75}\rely{0}{4}\rely{0}{46} };
            \node[right=.1cm of 3] (4) {\rely{1}{48}\rely{0}{6}\rely{0}{32}\rely{1}{29} };
            \node[right=.1cm of 4] (5) {\rely{1}{14}\rely{0}{41}\rely{1}{42}\rely{0}{23}};
            \node[right=.1cm of 5] (6) {\rely{1}{38}\rely{0}{55}\rely{1}{66}\rely{0}{38}};
            \node[right=.1cm of 6] (7) {\rely{1}{40}\rely{0}{38}\rely{0}{22}\rely{1}{74}};
            \node[right=.1cm of 7] (8) {\rely{0}{54}\rely{1}{54}\rely{0}{63}\rely{1}{66}};
            \node[right=.1cm of 8] (9) {\rely{0}{49}\rely{1}{28}\rely{1}{17}\rely{0}{80}};
            \node[right=.3cm of 9] (10) {};
        \end{tikzpicture}
        }

    \end{subfigure}
    \begin{subfigure}[t]{\columnwidth}
    \resizebox {\columnwidth} {!} {
    \begin{tikzpicture}
      \node (0) at (0,0) {};
      \node[right=.6cm of 0] (1) {c)};
      \node[right=.2cm of 1] (2) {\rely{0}{80}\rely{1}{75}\rely{1}{74}\rely{1}{66}};
      \node[right=.1cm of 2] (3) {\rely{1}{66}\rely{0}{63}\rely{0}{55}\rely{0}{54}};
      \node[right=.1cm of 3] (4) {\rely{1}{54}\rely{0}{53}\rely{0}{49}\rely{1}{48}};
      \node[right=.1cm of 4] (5) {\rely{0}{46}\rely{1}{42}\rely{0}{41}\rely{1}{40}};
      \node[right=.1cm of 5] (6) {\rely{1}{38}\rely{0}{38}\rely{0}{38}\rely{0}{32}};
      \node[right=.1cm of 6] (7) {\rely{1}{29}\rely{1}{28}\rely{0}{23}\rely{0}{22}};
      \node[right=.1cm of 7] (8) {\unrely{1}{18}\unrely{0}{18}\unrely{1}{17}\unrely{1}{17}};
      \node[right=.1cm of 8] (9) {\unrely{1}{14}\unrely{0}{6}\unrely{0}{4}\unrely{0}{3}};
      \node[right=.3cm of 9] (10) {};
    \end{tikzpicture}
    }
    \end{subfigure}
    \end{subfigure}
    \begin{subfigure}[t]{\columnwidth}
        \centering
        \begin{tikzpicture}
          \node (start) at (0,0) {}; 
            \node[left=0cm of start] (waist) {$\boldsymbol{f}_{\mathit{waist}}$};
            \node[right=8cm of waist] (forearm) {$\boldsymbol{f}_{\mathit{forearm}}$};

            \setlength{\baselineskip}{12pt}
            
            \begin{scope}[every node/.style={midway}]
            
            \draw[|-|](waist)--(forearm) node [sloped,above=1pt]{\SI{80}{\percent} similarity};
          
          \end{scope}
        \end{tikzpicture}
    \end{subfigure}
    \caption{Independent fingerprint generation on sensor positions waist and forearm: The graph shows energy levels above the average gait cycle $\boldsymbol{A}$ as blue and below as red areas. After quantization in (a), reliabilities are calculated and assigned to each bit in (b). The darker the color, the more reliable the bit. Finally, in (c) the fingerprint is sorted by reliability vector of the forearm with a cutoff at 24.}
    \label{fig:quantization}
\end{figure*}

\section{BANDANA}
\label{sec:pairing}
After recording raw accelerometer as well as gyroscope data and correcting sensor-orientations together with applying a band-pass filter, the gait cycle detection algorithm produces a periodic filtered signal.
For BANDANA's device-to-device authentication, shared secrets need to be generated based on these signals independently on participating devices and, in particular, without disclosing information on the gait sequence on the communication channel.

\subsection{Quantization}
To generate binary fingerprints from the continuous gait sequence, we propose a quantization algorithm inspired by \citeauthor{hoang2015gaitauthentication}~\cite{hoang2015gaitauthentication}.
Recall the definition of $Z_i$ from equation~(\ref{equatonZi}) with the normalized gait cycle $|Z_i|=\rho$ and $Z_i=\{Z_{i1},\dots,Z_{i\rho}\}$. 
We define the average gait cycle as
\begin{eqnarray}
\boldsymbol{A} &=& \boldsymbol{(}A_1, \dots , A_j , \dots A_{\rho}\boldsymbol{)}; \nonumber\\
A_j &=& \frac{\sum_{i=1}^{q} Z_{ij}}{q} \text{.} \nonumber
\end{eqnarray}
Fingerprint bits are extracted by calculating the energy difference between each gait cycle $Z_i$ and $\boldsymbol{A}$ as depicted in Figure~\ref{fig:quantization}.
To extract $b$ bit per $Z_i$, each $Z_i$ is split into $b$ parts of the same length $\rho / b$.
Thus, a binary fingerprint is defined by
\begin{eqnarray}
\boldsymbol{\tilde{f}} &=& \boldsymbol{(}\tilde{f}_{11}, \dots , \tilde{f}_{1\frac{\rho}{b}}, \dots, \tilde{f}_{b1}, \dots , \tilde{f}_{b\frac{\rho}{b}}\boldsymbol{)}\nonumber\\
  \tilde{f}_{ij}&=&\left\{\begin{array}{rl}
    1, & \delta_{ij} > 0\\
    0, &\mbox{otherwise.}
  \end{array}\right. \nonumber\\
  \delta_{ij} &=& \sum_{k=0}^{\rho / b} A_{i+k} - Z_{i+k, j} \text{.}\nonumber
\end{eqnarray}
as exemplary shown in Figure~\ref{fig:quantization} (a).
In the following, the fingerprint vector is written as
\begin{equation*}
\boldsymbol{\tilde{f}} = \boldsymbol{(}\tilde{f}_{1}, \dots , \tilde{f}_{M}\boldsymbol{)} \text{.}
\end{equation*}

\subsection{Reliability}
To calculate the reliability of the extracted bits, the differences of the quantization algorithm are stored as
\begin{equation*}
\boldsymbol{\delta} = \boldsymbol{(}\delta_{11}, \dots , \delta_{1b}, \dots, \delta_{q1}, \dots , \delta_{qb}\boldsymbol{)} \text{.}
\end{equation*}
The indices of $\boldsymbol{\delta}$ are sorted in descending order by the absolute value of each associated difference $| \delta_{ij} |$ to retrieve the reliability ordering
\begin{equation*}
\boldsymbol{r} = \boldsymbol{(}r_{1}, \dots, r_{M}\boldsymbol{)} \mbox{ with } r_{i}\geq r_{i+1} \text{.}
\end{equation*}
We will refer to $\boldsymbol{r}$ in the following as the \textit{reliability vector}.
It contains those indices which experienced the highest difference between the mean gait cycle $\boldsymbol{A}$ an an instantaneous normalized gait cycle $Z_j$.
These bits are most reliable in the sense that they have high probability to be identical for devices at arbitrary body positions. 
In Figure~\ref{fig:quantization} (b) we used colors to indicate the associated reliability.
The elements of $\boldsymbol{\tilde{f}}$ are then sorted according to their corresponding values of $\boldsymbol{r}$ and the most reliable first $N$ constitute the 
final fingerprint~$\boldsymbol{f} =\boldsymbol{(}f_{r_1},\dots,f_{r_N}\boldsymbol{)}$ (cf. Figure~\ref{fig:quantization} (c)).

\subsection{Fuzzy Cryptography}
To derive unique shared secrets on both participating devices without exchanging additional information for comparison, error correcting codes are utilized.
Error correcting codes are normally used to encode messages from the messagespace $m \in \mathcal{M}$ into codewords of the (larger) codespace $c \in \mathcal{C}$ introducing redundancies.
\begin{equation*}
     m \xrightarrow{Encode} c \text{.}
\end{equation*}
This process allows to correct errors introduced when transmitting $c$ over a lossy channel before decoding it back to $m$ with
\begin{equation*}
     c \xrightarrow{Decode} m \text{.}
\end{equation*}

We apply error correcting codes in a different way.
In a sense, our fingerprints $\boldsymbol{f}$ are lossy as they are not entirely equal on the devices trying to mutually authenticate.
Here, the codespace $\mathcal{C}$ is chosen in a way that we can directly pick a fingerprint $\boldsymbol{f}$ from this codespace and apply the $\mathit{Decode}$-method with
\begin{equation*}
     \boldsymbol{f} \xrightarrow{Decode} \boldsymbol{k}
\end{equation*}
to derive a binary key $\boldsymbol{k}$ that is error corrected.
Due to the usage of binary fingerprints we propose the usage of BCH codes over the Galois field $\mathbb{F}_2$.
A BCH code can be parameterized to correct up to $t$ errors, which in our case must be chosen carefully to allow for errors within different positions on the same body but not for correction of errors between different bodies.
As with the other parameters, $t$ is chosen based on our evaluation in Section~\ref{sec:evaluation}.

\subsection{Protocol}
Finally, we introduce BANDANA's full protocol flow between two devices $A$ and $B$ worn on the same body.
Following Figure~\ref{fig:flow}, the gait cycle detection is applied on recorded accelerometer data corrected by Madgwick's algorithm and Type-II Chebyshev bandpass filter.
For two co-aligned devices A (Alice) and B (Bob), fingerprints $\boldsymbol{f_A}$ ($\boldsymbol{f_B}$) and reliability vectors $\boldsymbol{r_A}$ ($\boldsymbol{r_B}$) are derived on both devices independently.
To utilize the same vector for reliability ordering on both sides, a random value $x_A$ ($x_B$) is transmitted together with $\boldsymbol{r_A}$ ($\boldsymbol{r_B}$).
The same reliability vector $\boldsymbol{r}$ is then used to sort $\boldsymbol{r_A}$ and $\boldsymbol{r_B}$.
To account for errors, we apply the BCH decoding-method to reduce both $\boldsymbol{r_A}$ and $\boldsymbol{r_B}$ to a unique $\boldsymbol{k}$, which is then used as the password for a Password-Authenticated Key Agreement~(PAKE).
Both devices now share the same secret~$\boldsymbol{s}$ protected by a key agreement authenticated by their gait fingerprints.
We propose the usage of a modern non-patented PAKE that feature additional countermeasures for low entropy passwords, such as Password Authenticated Key Exchange by Juggling~(J-PAKE)~\cite{Hao2010jpake} or Secure Remote Password protocol~(SRP)~\cite{wu1998srp}.

For devices with high clock drift, the protocol can be extended to allow for multiple tries with different fingerprints.
For this to work, before quantization the gait sequence should be shifted by \emph{half}-gait cycles to the right by both devices $A$ and $B$.

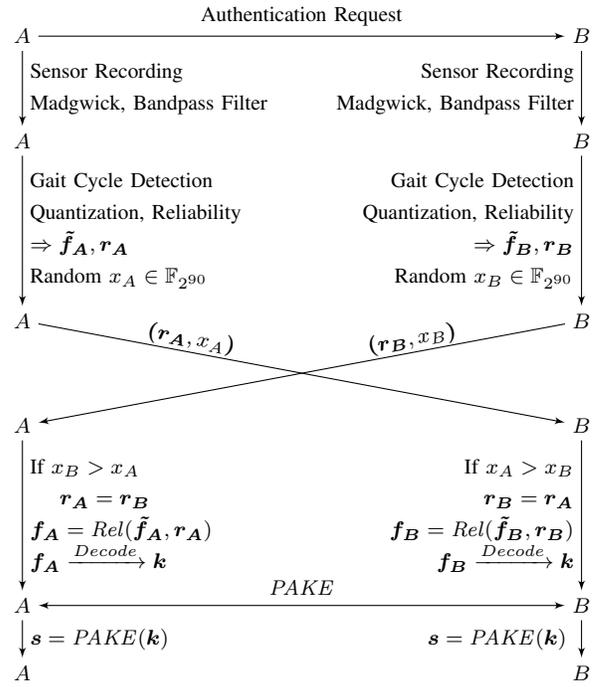
\begin{figure}
\centering
{\footnotesize
\begin{tikzpicture}
  \node (A0) at (0,0) {$A$};
  \node[below=1cm of A0] (A1) {$A$};
  \node[below=2cm of A1] (A2) {$A$};
  \node[below=1cm of A2] (A3) {$A$};
  \node[below=2cm of A3] (A4) {$A$};
  \node[below=.5cm of A4] (A5) {$A$};
  
  \node[right=7cm of A0] (B0) {$B$};
  \node[below=1cm of B0] (B1) {$B$};
  \node[below=2cm of B1] (B2) {$B$};
  \node[below=1cm of B2] (B3) {$B$};
  \node[below=2cm of B3] (B4) {$B$};
  \node[below=.5cm of B4] (B5) {$B$};
  
  \setlength{\baselineskip}{12pt}

  \begin{scope}[every node/.style={midway},>=latex']
  \draw[->]        (A0)--(B0) node [sloped,above=1pt] {
    Authentication Request
  };
  \draw[->, align=left]        (A0)--(A1) node [right] {
    Sensor Recording\\
    Madgwick, Bandpass Filter
  };
  \draw[->, align=left]        (A1)--(A2) node [right] {
    Gait Cycle Detection \\
    Quantization, Reliability\\
    $\Rightarrow \boldsymbol{\tilde{f}_A}, \boldsymbol{r_A}$\\
    Random $x_A \in \mathbb{F}_{2^{90}}$
  };
  \draw[->, align=right]        (B0)--(B1) node [left] {
    Sensor Recording\\
    Madgwick, Bandpass Filter
  };
  \draw[->, align=right]        (B1)--(B2) node [left] {
    Gait Cycle Detection \\
    Quantization, Reliability\\
    $\Rightarrow \boldsymbol{\tilde{f}_B}, \boldsymbol{r_B}$\\
    Random $x_B \in \mathbb{F}_{2^{90}}$
  };
  \draw[->]        (A2)--(B3) node [sloped,above=5pt,left] {
    \hspace*{-3cm}$\boldsymbol{(} \boldsymbol{r_A}, x_A \boldsymbol{)}$
  } node [sloped,below] {};
  \draw[->]        (B2)--(A3) node [sloped,above=5pt,right] {
    $\boldsymbol{(} \boldsymbol{r_B}, x_B \boldsymbol{)}$\hspace*{-3cm}
  } node [sloped,below] {};
  \draw[->, align=left]        (A3)--(A4) node [right] {
    If $x_B > x_A$\\
    ~~~~$\boldsymbol{r_A} = \boldsymbol{r_B}$\\
    $\boldsymbol{f_A} = \mathit{Rel}(\boldsymbol{\tilde{f}_A}, \boldsymbol{r_A})$\\
    $\boldsymbol{f_A} \xrightarrow{Decode} \boldsymbol{k}$
  };
  \draw[->, align=right]        (B3)--(B4) node [left] {
    If $x_A > x_B$\\
    ~~~~$\boldsymbol{r_B} = \boldsymbol{r_A}$\\
    $\boldsymbol{f_B} = \mathit{Rel}(\boldsymbol{\tilde{f}_B}, \boldsymbol{r_B})$\\
    $\boldsymbol{f_B} \xrightarrow{Decode} \boldsymbol{k}$
  };
  \draw[<->]        (A4)--(B4) node [sloped,above=1pt] {
    $\mathit{PAKE}$
  };
  \draw[->, align=left]        (A4)--(A5) node [right] {
    $\boldsymbol{s} = \mathit{PAKE}(\boldsymbol{k})$
  };
  \draw[->, align=right]        (B4)--(B5) node [left] {
    $\boldsymbol{s} = \mathit{PAKE}(\boldsymbol{k})$
  };

  \end{scope}
\end{tikzpicture} 
}
\caption{BANDANA protocol sequence between two devices $A$ and $B$ worn on the same body.}
\label{fig:flow}
\end{figure}

\section{Evaluation}\label{sec:evaluation}
In this section, we present the evaluation of our approach as well as experiments that led us to choose certain parameter configurations.

\subsection{Signal Coherence}
\begin{figure} 
    \centering%
    \setlength\figureheight{6cm}
    \setlength\figurewidth{\columnwidth}
    \input{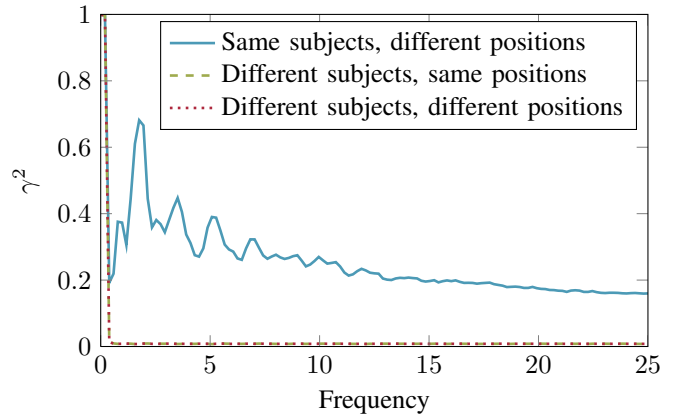}
    \caption{Average spectral coherence over full sensor readings of the Mannheim dataset for same and different subject.}
    \label{fig:coherence}
\end{figure}

After applying Madgwick's algorithm (cf. section~\ref{sec:data_prep}), we end up with sensor readings where the z-axis has been aligned to point to the ground.
This allows to examine the relation between these sensor readings.
For this, we calculate the spectral coherence for different sensor combinations to test whether any causality between readings taken simultaneously by sensors located in different positions on the same body exists --- apart from just the correlation for the respective motion in general.
Figure~\ref{fig:coherence} clearly shows that there is hardly any similarity between arbitrary recordings, but a considerably high correlation between those records being taken simultaneously.
Only between \SI{0}{\hertz} up to \SI{0.5}{\hertz}, a correlation between arbitrary records exists.
This leaves us with two major results:
\begin{itemize}
    \item There is a measurable causality between sensor readings taken simultaneously on the same body
    \item Some unwanted correlation at lower frequencies still exists
\end{itemize}

\subsection{Bandpass Filter}
\label{sec:bandpass}
Following the results from the previous section, we continue filtering the sensor readings.
As visualized in Figure~\ref{fig:coherence}, there still exists some unexpected correlation between arbitrary readings on low frequencies.
As these frequencies - up to approximately \SI{0.5}{\hertz} - only add noise, we would like to filter them out while keeping all the frequencies above.
We thus employ a Type-II Chebyshev filter, which is known to have a very steep drop at the cutoff frequency.
Furthermore, in contrast to Type-I, Type-II Chebyshev filters do not have any ripple in the passband.

Researchers in the domain of Activity Recognition report that human motion does not affect any frequencies significantly above \SI{10}{\hertz}~\cite{lester2004you}.
Based on this observation and the coherence depicted in Figure~\ref{fig:coherence}, we decided to choose an upper cutoff frequency of \SI{12}{\hertz}.

\subsection{Reliability}
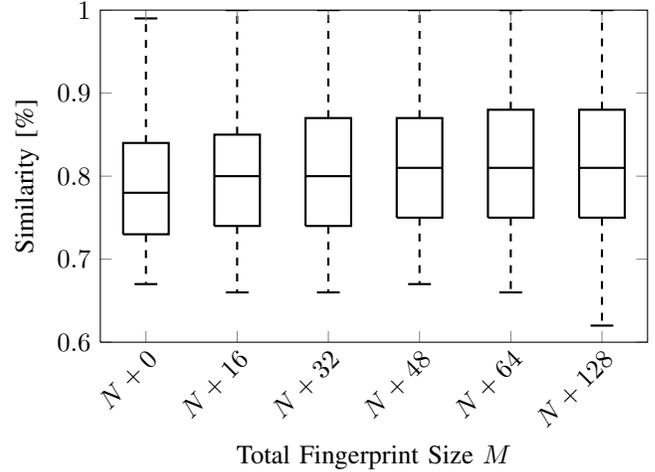
\begin{figure}
    \centering%
    \setlength\figureheight{6cm}
    \setlength\figurewidth{\columnwidth}
\begin{tikzpicture}

\begin{axis}[
xlabel={Total Fingerprint Size $M$},
ylabel={Similarity [\%]},
xmin=0.5, xmax=6.5,
ymin=0.6, ymax=1,
axis on top,
width=\figurewidth,
height=\figureheight,
xtick={1,2,3,4,5,6},
xticklabels={$N+0$,$N+16$,$N+32$,$N+48$,$N+64$,$N+128$},
x tick label style={rotate=45, anchor=north east, inner sep=.5mm} 
]
\addplot [thick, black]
table {%
0.75 0.73
1.25 0.73
1.25 0.84
0.75 0.84
0.75 0.73
};
\addplot [thick, black, dashed]
table {%
1 0.73
1 0.67
};
\addplot [thick, black, dashed]
table {%
1 0.84
1 0.99
};
\addplot [thick, black]
table {%
0.875 0.67
1.125 0.67
};
\addplot [thick, black]
table {%
0.875 0.99
1.125 0.99
};
\addplot [thick, black]
table {%
0.75 0.78
1.25 0.78
};
\addplot [thick, black]
table {%
1.75 0.74
2.25 0.74
2.25 0.85
1.75 0.85
1.75 0.74
};
\addplot [thick, black, dashed]
table {%
2 0.74
2 0.66
};
\addplot [thick, black, dashed]
table {%
2 0.85
2 1
};
\addplot [thick, black]
table {%
1.875 0.66
2.125 0.66
};
\addplot [thick, black]
table {%
1.875 1
2.125 1
};
\addplot [thick, black]
table {%
1.75 0.8
2.25 0.8
};
\addplot [thick, black]
table {%
2.75 0.74
3.25 0.74
3.25 0.87
2.75 0.87
2.75 0.74
};
\addplot [thick, black, dashed]
table {%
3 0.74
3 0.66
};
\addplot [thick, black, dashed]
table {%
3 0.87
3 1
};
\addplot [thick, black]
table {%
2.875 0.66
3.125 0.66
};
\addplot [thick, black]
table {%
2.875 1
3.125 1
};
\addplot [thick, black]
table {%
2.75 0.8
3.25 0.8
};
\addplot [thick, black]
table {%
3.75 0.75
4.25 0.75
4.25 0.87
3.75 0.87
3.75 0.75
};
\addplot [thick, black, dashed]
table {%
4 0.75
4 0.67
};
\addplot [thick, black, dashed]
table {%
4 0.87
4 1
};
\addplot [thick, black]
table {%
3.875 0.67
4.125 0.67
};
\addplot [thick, black]
table {%
3.875 1
4.125 1
};
\addplot [thick, black]
table {%
3.75 0.81
4.25 0.81
};
\addplot [thick, black]
table {%
4.75 0.75
5.25 0.75
5.25 0.88
4.75 0.88
4.75 0.75
};
\addplot [thick, black, dashed]
table {%
5 0.75
5 0.66
};
\addplot [thick, black, dashed]
table {%
5 0.88
5 1
};
\addplot [thick, black]
table {%
4.875 0.66
5.125 0.66
};
\addplot [thick, black]
table {%
4.875 1
5.125 1
};
\addplot [thick, black]
table {%
4.75 0.81
5.25 0.81
};
\addplot [thick, black]
table {%
5.75 0.75
6.25 0.75
6.25 0.88
5.75 0.88
5.75 0.75
};
\addplot [thick, black, dashed]
table {%
6 0.75
6 0.62
};
\addplot [thick, black, dashed]
table {%
6 0.88
6 1
};
\addplot [thick, black]
table {%
5.875 0.62
6.125 0.62
};
\addplot [thick, black]
table {%
5.875 1
6.125 1
};
\addplot [thick, black]
table {%
5.75 0.81
6.25 0.81
};
\end{axis}

\end{tikzpicture}
    \caption{Different fingerprint sizes $M$ with cutoff at $N=128$ to evaluate the influence of $\textit{Rel}()$ on intra-body similarity. Each value in this graph is defined by the similarity of two \emph{differing} sensor positions within the same subject (intra-body). The graph contains all possible similarities within each subject and all her sensor position-combinations. Fingerprints are generated by a sliding window over the sensor data with half-overlapping windows. Only fingerprints generated from the same windows are matched against each other.}
    \label{fig:reliability}
\end{figure}

Our quantization scheme defines that iff $\delta_{ij} > 0$ for fixed $i,j$ is true for Alice, the same has to apply for Bob for at least \SI{80}{\percent}.
Some $Z_{ij}$ are less prone to leading to different bits between sensors at different body locations than others, namely those with a higher difference $\delta_{ij}$ to the mean gait $\boldsymbol{A}$.
Both Alice and Bob keep a reliability value for each bit of the fingerprint.
According to the BANDANA-Protocol as shown in Figure~\ref{fig:flow}, one of these reliability vectors is chosen randomly and the fingerprint is sorted by each party following tis reliability vector's order of indices (see Figure~\ref{fig:quantization}).
In a last step, the fingerprint's most unreliable bits are discarded.
To show the viability of this approach, we calculated the fingerprints' similarity over all 15 subjects and all 7 sensor positions.
As shown in Figure~\ref{fig:reliability}, we chose different fingerprint sizes $M$ with cutoff at $N=128$ to test how many additional bits should be discarded by BANDANA to gain the best similarity.
The mean-similarity improves with greater values of $M$ and settles around $N+64$ with an average improvement of approximately 4\%.
Thus, we chose $N+64$ for our configuration.

\subsection{Discriminability of Intra- and Inter-body Fingerprints}
\label{sec:discriminability}
\begin{figure}
    \centering%
    \setlength\figureheight{6cm}
    \setlength\figurewidth{\columnwidth}
    \input{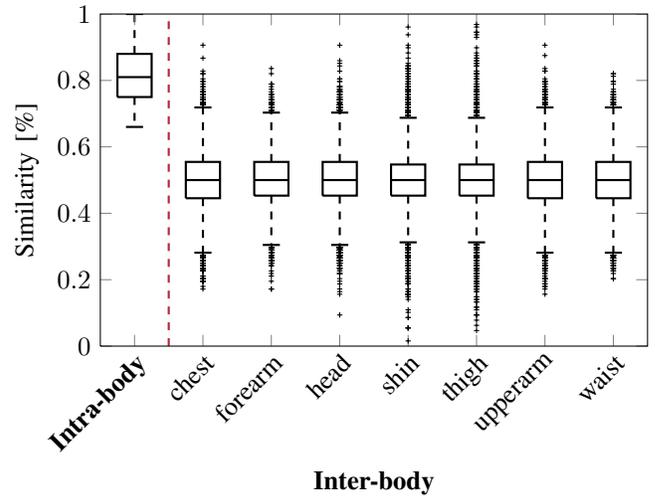}
    \caption{Comparison of intra-body against inter-body similarity. Each value in the \emph{intra-body} boxplot is defined by the similarity of two \emph{different} sensor positions within the same subject (all possible similarities within each subject and all her sensor position-combinations). For the \emph{inter-body} test, each boxplot defines a different sensor position. Only \emph{different} subjects are tested against each other with the \emph{same} sensor positions. Fingerprints are generated by a sliding window over the sensor data with half-overlapping windows for $M=192$ with cutoff at $N=128$.}
    \label{fig:intra-inter}
\end{figure}

Figure~\ref{fig:intra-inter} illustrates the discriminability between intra-body and inter-body fingerprints.
The intra-body case with $M=N+64$ with $N=128$ is compared against the inter-body cases of different sensor positions.
Here, we evaluated whether similarities exist between different subjects but same sensor positions.
While the intra-body case tests only similarities between differing sensor position on the same body (315 similarities), the inter-body case is much larger.
Each inter-body position case contains 8880300 similarities.
As expected, the mean similarity between different subjects is \SI{50}{\percent}.
It is important to note that this test evaluates the worst case of brute forcing all possible combinations between subjects.
In reality, an attacker is constrained to ${\sim432}$ tries per day (cf. Section~\ref{sec:attackmodel}).
In the inter-body case, it can be seen that a small number of fingerprints match with unexpected high similarity values (outliers).
We assume that these collisions happen in case of gait sequences with very low entropy still exhibiting specific pattern due to the design of the quantization scheme.
While this should be investigated further, only \SI{0.0642}{\percent} of these collisions show similarity values above \SI{80}{\percent}.

\subsection{Similarities between Sensor Position-Combinations}
Table~\ref{tab:sensor-positions} illustrates how well differing sensor positions authenticate against each other.
We found out that chest against other positions and head against other positions perform worse while forearm and waist perform best.

\ctable[
    caption = {Detailed comparison of sensor position-combinations worn on the same body (intra-body). Shown is the mean over all 15 subjects.},
    label = tab:sensor-positions,
    pos = tbp,
]{lllllllll}{%
}{
 & \roth{chest} & \roth{forearm} & \roth{head} & \roth{shin} & \roth{thigh} & \roth{upperarm} & \roth{waist}  \ML
chest &  1.0 & 0.82 & 0.74 & 0.78 & 0.78 & 0.88 & 0.81  \NN
forearm &  0.82 & 1.0 & 0.8 & 0.81 & 0.88 & 0.89 & 0.89  \NN
head &  0.74 & 0.8 & 1.0 & 0.8 & 0.76 & 0.77 & 0.78  \NN
shin &  0.78 & 0.81 & 0.8 & 1.0 & 0.77 & 0.78 & 0.8  \NN
thigh &  0.78 & 0.88 & 0.76 & 0.77 & 1.0 & 0.85 & 0.84  \NN
upperarm &  0.88 & 0.89 & 0.77 & 0.78 & 0.85 & 1.0 & 0.88  \NN
waist &  0.81 & 0.89 & 0.78 & 0.8 & 0.84 & 0.88 & 1.0  \LL
}

\subsection{Statistical Bias}
\begin{figure}
 \centering%
 \includegraphics[width=0.5\columnwidth]{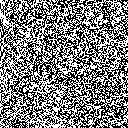}
 \caption{Illustration of 128 binary keys after removing 64 unreliable bits from the fingerprint. Each row contains one 128 bit fingerprint with $1=\text{black}$ and $0=\text{white}$.}
 \label{fig:random}
\end{figure}

For the robustness against a potent adversary, it is important that the keys generated from gait sequences are random. 
For instance, Figure~\ref{fig:random} exemplarily depicts 128 keys we extracted using BANDANA with fingerprint length $M=192$ bits after removing $64$ unreliable bits (cutoff at $N=128$ bits) for an intuitive illustration of the randomness of the generated fingerprints.

We rigorously tested the keys generated by BANDANA against statistical bias and employed the dieHarder battery of statistical tests for this end~\cite{statistical_Brown_0000}.
While these tests can not replace cryptanalysis they are designed to uncover bias and dependency in the pseudo random sequence. 
Every test has an expected distribution of outcomes.
Test runs produce a value that is compared to the theoretical outcome. 
A p-value, describing the probability that a real Random Number Generator (RNG) would produce this outcome, between 0 and 1 is computed. 
A good RNG will have a range of p-values that follows a uniform distribution.
A p-value below a fixed significance level $\alpha = 0.001$ indicates a failure of the PRNG with probability $1 - \alpha$.
For instance, a p-value $\leq0.05$ is exptected $5$\% of the time. 

%



\begin{figure*}
\includegraphics[width=\textwidth]{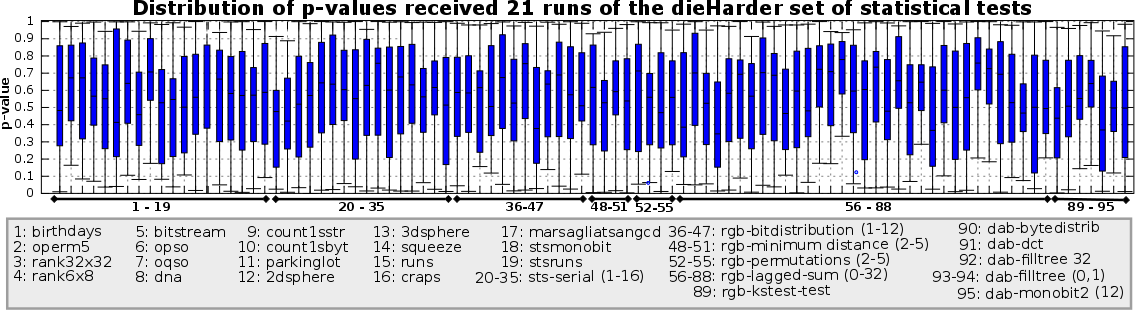}
\caption{Distribution of p-values achieved for $128$bit keys (fingerprint length $M=192$, $64$ unreliable bits removed) in the various statistical tests of the dieHarder set of statistical tests.}
\label{fig:dieHarder}
\end{figure*}
Our results are depicted in Figure~\ref{fig:dieHarder}.
Observe that the p-values are well distributed over the complete range and clustered in the center which indicates a good random distribution of the p-values.





\subsection{Final Parameters}
Throughout the paper, we introduced parameters without assigning a definitive value.
We now present the configuration we propose for a deployment of BANDANA in real-world applications.
The parameters have been carefully chosen based on our evaluations performed on the datasets from Mannheim and Osaka University.
When using an accelerometer resolution of \SI{50}{\hertz}, we propose a resampling rate of $\rho = \SI{40}{}$ for bit extraction of $b = 4$ bits per gait cycle $R_i$ resulting in $\tau = \rho / b = 10$.
We target $M=192$\;bit gait fingerprints with a cutoff at $N=128$ bit, i.e., for $b = 4$ we extract $q = 48$ gait cycles.
An upper bound for the required length of the data $\boldsymbol{r}$ is given with $48 \cdot {\sim}2\;s=96\;s$.

Following the results depicted in Figure~\ref{fig:intra-inter}, we chose to parameterize the BCH codes to allow correction of at maximum \SI{20}{\percent} of the bits in the fingerprint with $t= \left\lfloor 128 \cdot 0.2  \right\rfloor = 25$.
Consequently, at least \SI{80}{\percent} similarity between the fingerprints is required.
This results in a $103$-bit security level for $\boldsymbol{k}$ used as a password for PAKE.
While we have shown that the entropy of our fingerprints is sufficiently high, PAKE takes additional countermeasures for low entropy.

\section{Security Model}
\label{sec:attackmodel}
In the following, we analyze BANDANA's security model by discussing possible attack scenarios.
The attacks are discussed in order of increasing severity.

\subsubsection{Mimic Gait}
An imposter could try to mimic the gait of the victim to produce fingerprints above \SI{80}{\percent} similarity.
In BANDANA, the impersonator is constrained to mimic the gait for a very specific time frame where the authentication between the impersonator's device and the victim's device happen.
While we do not have actual sensor data of impersonators, the evaluation of the Mannheim dataset show that there are only a small number of accidental collisions between fingerprints from different subjects (cf. Section~\ref{sec:discriminability}) over all possible pairwise comparisons of all 14:1 subjects at 10-12 minutes duration each.
As shown in \cite{gafurov2007spoof}, a minimal-effort impersonation attack does not improve the chances of success.
In BANDANA, the default configuration allows for one try of PAKE only, before starting a completely new authentication process.
For $M=192$ bit long fingerprints, BANDANA's full process takes up to ${\sim\SI{200}{\second}}$.
Thus, an optimal imposter --- one who is always following the victim --- is constrained to ${\sim432}$ tries per day.

\subsubsection{Brute Force} 
Without requiring additional knowledge about the victim's gait, an attacker may want to brute force keys $\boldsymbol{k}$ by exhausting the space of all possible keys $\mathcal{C}=\mathbb{F}_{2^{103}}$.
As discussed in the previous scenario, an attacker is constrained to ${\sim432}$ tries per day, where every try generates a new $\boldsymbol{k}$ completely independent from the previous one.
Thus, previously tried keys need to be tried again making it impossible to exhaust $\mathcal{C}$.

\subsubsection{Video Recording}
An attacker controlling surveillance cameras could create a video recording of the victim's gait for the timespan where the device-to-device authentication happens.
Using motion detection software she could try to infer acceleration values to generate similar fingerprint.
It will be difficult to extract all acceleration changes by just extracting features from a video.
Recall that our fingerprint is based on tiny convulsions between a sliding window mean gait $\boldsymbol{A}$ and the instantaneous gait cycle $Z_i$, happening on the z-axis pointing towards the ground.
For every gait cycle $b=4$ bits are extracted from these changes.
While we believe that it will be difficult to accomplish this attack, a detailed analysis is left for future work. 

\subsubsection{Attach Malicious Device}
To access the BAN, an attacker could attach a malicious device to the body of the victim, e.g. by slipping a small sensor node into the victim's jacket or by selling a compromised device to the victim.
This device could create a second communication channel to forward traffic from inside the BAN to an outsider.
Due to the fact that BANDANA works without explicit user interaction, this attack could succeed if executed properly and unnoticed.
We would like to remark, though, that this physical attack also contains significant risk for the attacker to be revealed when such malicious device is detected.

\section{Conclusion}\label{sec:conclusion}
We have presented BANDANA: the first-ever implicit secure device-to-device authentication scheme for devices worn on the same body.
No user interaction is required to establish shared secrets implicitly and authenticated by fingerprints generated from the user's gait.
The protocol accounts for errors without comparing the fingerprints directly, but utilizes fuzzy cryptography based on error correcting codes.
A novel quantization method for independently generating similar fingerprints at differing sensor positions has been proposed and evaluated.
By selecting only reliable bits, we were able to boost the similarity by $\SI{4}{\percent}$.
Our final fingerprints between devices worn on the same body have a similarity of $\SI{82}{\percent}$ in comparison to devices worn on different bodies with $\SI{50}{\percent}$.
Therefore, we were able to show that BANDANA works considering its security model.
Our main future work will be the investigation of the small number of matching fingerprints in the inter-body case and on a detailed entropy analysis.


\renewcommand*{\bibfont}{\small}
\sloppy
\printbibliography

\end{document}